\documentclass[proceedings, preprint]{rmaa}

\usepackage{paralist}

\usepackage{psfrag,color}

\SetYear{2007} \SetConfTitle{XII Latin American Regional IAU
Meeting}

\title{Neutron Stars and their Magnetic Fields}

\author{
  Andreas Reisenegger\altaffilmark{1}
}

\altaffiltext{1}{Departamento de Astronom{\'\i}a y
Astrof{\'\i}sica, Pontificia Universidad Cat\'olica de Chile,
Casilla 306, Santiago 22, Chile (areisene@astro.puc.cl).}

\shortauthor{Reisenegger}
\shorttitle{Neutron Stars}

\listofauthors{A. Reisenegger}
\indexauthor{Reisenegger, A.}

\abstract{Neutron stars have the strongest magnetic fields known
anywhere in the Universe. In this review, I intend to give a
pedagogical discussion of some of the related physics. Neutron
stars exist because of Pauli's exclusion principle, in two senses:
1) It makes it difficult to squeeze particles too close together,
in this way allowing a mechanical equilibrium state in the
presence of extremely strong gravity. 2) The occupation of
low-energy proton and electron states makes it impossible for
low-energy neutrons to beta decay. A corollary of the second
statement is that charged particles are necessarily present inside
a neutron star, allowing currents to flow. Since these particles
are degenerate, they collide very little, and therefore make it
possible for the star to support strong, organized magnetic fields
over long times. These show themselves in pulsars and are the most
likely energy source for the high X-ray and gamma-ray luminosity
``magnetars''. I briefly discuss the possible origin of this field
and some physical constraints on its equilibrium configurations.}

\resumen{Las estrellas de neutrones poseen los campos magn\'eticos
m\'as intensos que conocemos. Aqu{\'\i} se pretende dar una
discusi\'on pedag\'ogica de parte de la f{\'\i}sica relacionada.
Las estrellas de neutrones existen gracias al principio de
exclusi\'on de Pauli, en dos sentidos: 1) \'Este hace dif{\'\i}cil
comprimir muchas part{\'\i}culas en un espacio peque\~no,
permitiendo un estado de equilibrio mec\'anico en presencia de una
fuerza de gravedad muy intensa. 2) La ocupaci\'on de estados
cu\'anticos de baja energ{\'\i}a por parte de protones y
electrones impide el decaimiento beta de neutrones poco
energ\'eticos. Un corolario de esto es que necesariamente hay
part{\'\i}culas cargadas en la estrella, permitiendo el flujo de
corrientes el\'ectricas. Como las part{\'\i}culas son degeneradas,
colisionan muy poco, haciendo posible la existencia prolongada de
campos magn\'eticos intensos. Estos se revelan en pulsares y son
la m\'as probable fuente de energ{\'\i}a para la alta luminosidad
en rayos X y gamma en ``magnetares''. Discuto brevemente el
posible origen de estos campos magn\'eticos, as{\'\i} como algunas
consideraciones f{\'\i}sicas que restringen a sus configuraciones
de equilibrio.}

\addkeyword{dense matter} \addkeyword{(magnetohydrodynamics:) MHD}
\addkeyword{magnetic fields} \addkeyword{stars: magnetic fields}
\addkeyword{stars: neutron} \addkeyword{stars: white dwarfs}

\begin{document}
\maketitle

\section{Introduction}
\label{sec:intro}

This text aims at giving a pedagogical introduction to the physics
of neutron stars and their magnetic fields, starting from
undergraduate physics (Quantum Mechanics and Electromagnetism) and
leading up to some current research questions such as the origin
and equilibrium configuration of the magnetic field and clarifying
some misconceptions appearing in the recent literature. All
equations used here can be derived ``on the back of an envelope'',
based only on undergraduate physics, and students are strongly
encouraged to do so.

In the philosophy of keeping the treatment simple and aiming at a
good understanding of basic questions rather than introducing all
the complications that might conceivably arise, I consider an
extremely conservative model of neutron stars that includes
neutrons, protons, and electrons, all treated as degenerate and
mostly non-interacting fermions. This neglects many extremely
interesting (but uncertain) issues such as neutron superfluidity,
proton superconductivity, or quark deconfinement and condensation.
For more comprehensive treatments of neutron star physics, also
addressing many of the ``exotic'' issues, I suggest the classic
book of \citet{Shapiro} and the recent volumes of
\citet{Glendenning} and \citet{Haensel}, as well as other works
mentioned in specific places of the text. For an inspiring popular
history of compact star research, written by one of its main
participants, see \citet{Thorne}.

\section{Degenerate fermions}
\label{sec:fermions}

The lowest energy state of a system of fermions is the ``Fermi
sea'' or momentum-space ``Fermi sphere'', in which $N$ fermions
occupy the least energetic orbitals available to them. If the
fermions are confined to a real-space volume $V$ and otherwise
non-interacting, they will agglomerate in momentum space in a
sphere around $\vec p=0$ with radius $p_F$ (the ``Fermi
momentum''), so the volume occupied in 6-dimensional phase space
$(x,y,z,p_x,p_y,p_z)$ is ${\cal V}_6=V\times(4\pi p_F^3/3)$. It
can be shown (e.~g., by solving the Schr\"odinger or Dirac
equation in a box) that there are $h^{-3}$ single-particle
orbitals per unit phase-space volume, where $h=2\pi\hbar$ is
Planck's constant. For spin-$1/2$ fermions such as electrons,
protons, or neutrons, Pauli's exclusion principle allows two
particles to be put in each orbital (with spin projections
$s_z=\pm {1/2}$ on an arbitrary axis), thus the total occupied
phase-space volume must also be equal to ${\cal V}_6=Nh^3/2$.
Equating both expressions for ${\cal V}_6$, one finds the Fermi
momentum as a function of the number density of fermions, $n\equiv
N/V$,
\begin{equation}\label{p_F}
p_F=\hbar(3\pi^2n)^{1/3},
\end{equation}
which holds regardless of how relativistic the particles are.

Global properties of the fermion system, such as its total energy
$E$, can be calculated as sums over all particles, i.~e.,
integrals over momentum space,
\begin{equation}\label{energy}
E={2V\over h^3}\int_{|\vec p|<p_F}d^3p~\varepsilon(\vec p),
\end{equation}
where one should generally use the relativistic energy-momentum
relation $\varepsilon(p)=[(mc^2)^2+(cp)^2]^{1/2}$, where $m$ is
the mass of the fermions and $c$ is the speed of light. These
integrals can be done analytically (see \citealt{Shapiro});
however, the calculation greatly simplifies in both the
non-relativistic limit ($p\ll mc$), in which
$\varepsilon(p)\approx mc^2+p^2/(2m)$, and in the extreme
relativistic limit ($p\gg mc$), in which $\varepsilon(p)\approx
cp$. In these limits, the pressure $P=-(\partial E/\partial V)_N$
takes a polytropic form $P\propto n^\gamma$, respectively
\begin{equation}\label{nonrel}
P\approx(3\pi^2)^{2\over 3}{\hbar^2\over 2m}~n^{5\over 3}
\end{equation}
and
\begin{equation}\label{xrel}
P\approx(3\pi^2)^{1\over 3}{\hbar c\over 4}~n^{4\over 3}.
\end{equation}
(For rough estimates, it is useful to remember that, aside from a
factor $0.4-0.25$, the pressure is just $n$ times the kinetic part
of the Fermi energy.)

\section{White dwarfs and neutron stars}
\label{sec:stars}

Since the temperature is essentially zero (compared to the Fermi
energies), the structure of degenerate stars is determined by only
two of the standard four equations of stellar structure, whose
Newtonian (weak-gravity) form is
\begin{equation}\label{structure}
{dP\over dr}=-{GM(r)\rho\over r^2},\qquad {dM(r)\over dr}=4\pi
r^2\rho,
\end{equation}
which can be combined to yield an order-of-magnitude expression
for the pressure required to sustain a star of given total mass
$M$ and characteristic mass density $\rho$ against its own
gravity,
\begin{equation}\label{hydropressure}
P\sim GM^{2/3}\rho^{4/3}.
\end{equation}

The ratio $\rho/n$ is generally constant, taking similar values:
\begin{itemize}
\item $\rho/n\approx Am_u/Z\approx 2m_u$ in white dwarfs, in which
most of the mass density is provided by ions of mass $Am_u$ (where
$m_u$ is the atomic mass unit) and charge $Ze$, while the
degeneracy pressure is due to the electrons, and \item
$\rho/n\approx m_n\approx m_u$ in neutron stars, where neutrons
(of mass $m_n$) dominate both the pressure and the mass density.
\end{itemize}

In the low-density, non-relativistic limit of eq.~(\ref{nonrel}),
the latter can be combined with eq.~(\ref{hydropressure}) to
estimate an equilibrium radius for the star,
\begin{equation}\label{radius}
R\sim\left(n\over\rho\right)^{5\over 3}{\hbar^2\over
Gm}M^{-{1\over 3}}.
\end{equation}
Due to the widely different fermion masses $m$ ($m_n\approx
1839~m_e$), the radii for white dwarfs and neutron stars are very
different ($\sim 10^4~\mathrm{km}$ for the former and $\sim
10~\mathrm{km}$ for the latter, at a similar, solar mass $M\sim
M_\odot$). Eq.~(\ref{radius}) shows that, as the mass increases,
the stars become smaller and denser and thus their matter more
relativistic. In the limit of high density, the hydrostatic
pressure of eq.~(\ref{hydropressure}) needs to be provided by
extremely relativistic particles (eq.~[\ref{xrel}]), yielding a
maximum mass
\begin{equation}\label{Mmax}
M_\mathrm{max}\sim\left(\hbar c\over G\right)^{3\over
2}\left(n\over\rho\right)^2\sim\left(m_\mathrm{Pl}\over
m_u\right)^3~m_u\sim M_\odot,
\end{equation}
where $m_\mathrm{Pl}=(\hbar c/G)^{1\over 2}$ is the ``Planck
mass'', a natural mass scale for quantum gravity. For white
dwarfs, this maximum mass is the well-known ``Chandrasekhar
limit'' $M_\mathrm{Chandra}\approx 1.4~M_\odot$, beyond which a
degenerate stellar core collapses to form a neutron star or black
hole. Note that the formation of a neutron star involves the
release of the binding energy $\sim GM^2/R\sim 0.2~Mc^2$ as
neutrinos (and perhaps gravitational waves), so a neutron star
mass can be lower than the Chandrasekhar limit by a corresponding
amount. In fact, a neutron star mass as low as $1.25~M_\odot$ was
measured in the recently discovered double pulsar system
\citep{Lyne}.

White dwarfs are well described by Newtonian gravity, and the main
inter-particle forces are electrostatic, for which the interaction
energies are much smaller than the kinetic energies, so the
physics described above suffices to obtain an accurate description
of the structure of a white dwarf. In neutron stars, such
estimates are generally not accurate, since at super-nuclear
densities the baryons (neutrons and protons) interact strongly
with each other (interaction energies comparable to the kinetic
energies), and the effects of General Relativity become important
(see below). Instructions for students to construct more realistic
neutron star models numerically have been given by
\citet{undergraduates1} and \citet{undergraduates2}.
Eq.~(\ref{Mmax}) suggests a maximum mass for neutron stars at most
a few times larger than $M_\mathrm{Chandra}$. This is likely to be
correct, but its precise value is not known, mainly due to the
uncertain strong interactions among the neutrons.

Given the basic structural parameters of the stars (here taken as
$M=M_\odot$ and $R=10~\mathrm{km}$), one can obtain several other
interesting quantities. The escape speed
$v_\mathrm{esc}=(2GM/R)^{1\over 2}=0.5~c$
confirms that neutron star gravity is ``strong'' and
general-relativistic effects can be important. Their minimum
allowed rotation period (at which the centrifugal force on the
equator equals the gravitational force)
$P_\mathrm{min}=2\pi[R^3/(GM)]^{1\over 2}=0.5~\mathrm{ms},$
is much shorter than for any other kind of stars (including white
dwarfs, for which $P_\mathrm{min}\sim 20~\mathrm{s}$), a crucial
argument in identifying pulsars as neutron stars.

Finally, one may estimate a safe upper bound on a typical magnetic
field $B$ in a neutron star by requiring that the magnetic energy
be lower than the gravitational binding energy (so the Lorentz
force does not exceed the gravitational force)\footnote{As
customary in theoretical astrophysics, I use Gaussian cgs units,
in which the unit of magnetic field is the gauss,
$\mathrm{G}\equiv\mathrm{erg}^{1\over 2}\mathrm{cm}^{-{3\over
2}}$.},
\begin{equation}\label{magenergy}
{B^2\over 8\pi}{4\pi R^3\over 3}<{GM^2\over R},
\end{equation}
yielding
\begin{equation}\label{Bmax}
B_\mathrm{max}\sim 10^{18}~\mathrm{G}.
\end{equation}

\section{Why {\it neutron} stars?}
\label{charged}

The name ``neutron stars'' incorrectly suggests stars composed
exclusively of neutrons. However, additional particles inside
these stars play a crucial role. A neutron ($n$) in vacuum decays
by the weak interaction process $n\to p+e+\bar\nu_e$ (beta decay)
into a proton ($p$), an electron ($e$), and an electron
antineutrino ($\bar\nu_e$), with a half-life close to 15 minutes.
This is impeded in very dense matter by the Pauli exclusion
principle: If all the low-energy proton and electron states are
already occupied, only suffiently energetic neutrons can decay. On
the other hand, if many protons and electrons are present, some of
these will be energetic enough to combine into neutrons by inverse
beta decay, $p+e\to n+\nu_e$, where $\nu_e$ stands for an electron
neutrino. In a neutron star, the neutrons and protons will be
confined by gravity, the electrons by the electrostatic potential
of the protons (e.~g., \citealt{R06}), while neutrinos and
antineutrinos are unbound and escape, contributing to the cooling
of the star (e.~g., \citealt{Yakovlev01}). Direct and inverse beta
decays are in balance if the chemical potentials\footnote{At zero
temperature, these chemical potentials reduce to the respective
Fermi energies.} of neutrons ($\mu_n$), protons ($\mu_p$), and
electrons ($\mu_e$) satisfy the relation $\mu_n=\mu_p+\mu_e$,
which forces the coexistence of a small fraction (few percent, but
density-dependent) of charged particles with a much larger number
of neutrons (e.~g., \citealt{Shapiro}). Additional particles (both
charged and uncharged) can appear by other weak decay processes at
densities higher than typical nuclear densities.

In addition to stabilizing the neutrons, the charged particles
play two important roles regarding the magnetic fields and their
evolution. Being charged, these particles can generate electrical
currents, which support potentially very strong magnetic fields.
In addition, since the proton fraction $Y\equiv n_p/n$ depends on
density ($n_i$ stands for the number density of particle species
$i=n,p,e$, and $n\equiv n_n+n_p$ is the total baryon density),
neutron star matter is inhomogeneous, stabilizing it with respect
to convective overturn \citep{Pethick92,RG92,R01a}. As discussed
below, this is likely to have an important stabilizing effect on
magnetic field configurations.

\section{Faraday's law and astrophysical magnetic fields}

Long-lived magnetic fields are ubiquitous in the Universe, and
neutron stars are no exception. These nearly static magnetic
fields must have currents acting as sources, according to
Amp\'ere's law,
\begin{equation}\label{Ampere}
\nabla\times\vec B={4\pi\over c}\vec j.
\end{equation}
Currents imply charges of one sign (i.~e., electrons) moving with
respect to those of the opposite sign (protons or other ions). On
time scales much shorter than the observed lifetimes of the
fields, these particles suffer Coulomb collisions, which would
damp their relative motion, if it were purely due to inertia.

Thus, in order for the magnetic fields to survive, a much larger
``inertia'' is required, which is provided by Faraday's induction
law: Any change of a magnetic field induces an electric field,
\begin{equation}\label{Faraday}
{\partial\vec B\over\partial t}=-c\nabla\times\vec E.
\end{equation}
The electric field keeps the current going in spite of the
frequent collisions, as described by Ohm's law,
\begin{equation}\label{Ohm}
{\vec j\over\sigma}=\vec E,
\end{equation}
where the conductivity $\sigma$ is inversely proportional to the
collision rate. Combining eqs. (\ref{Ampere}), (\ref{Faraday}),
and (\ref{Ohm}), one obtains a diffusion equation for the magnetic
field,
\begin{equation}\label{diffusion}
{\partial\vec B\over\partial t}=-\nabla\times\left({c^2\over
4\pi\sigma}\nabla\times\vec B\right),
\end{equation}
suggesting an Ohmic (or resistive) decay time $t_B\sim 4\pi\sigma
R^2/c^2$, where $R$ is a characteristic length scale, such as a
stellar radius. In astrophysical plasmas, the length scales are
enormously larger than in laboratory conditions, which allows
astrophysical magnetic fields to be so long-lived. In the
particular case of neutron stars, the Pauli principle also makes
it hard for particles to be scattered into different quantum
states, therefore enhancing the conductivity. Based on this,
\citet{Baym69} showed that neutron star magnetic fields might live
longer than a Hubble time.

The only ``loophole'' that allows for a significant evolution of
the magnetic field is to allow for a velocity field $\vec v_e$ of
the electrons in the reference frame of interest (i.~e., the
center-of-mass frame of a star). This velocity adds a magnetic
force term $(1/c)\vec v_e\times\vec B$ on the right-hand side of
eq. (\ref{Ohm}), and correspondingly a term $\nabla\times(\vec
v_e\times\vec B)$ on the right-hand side of eq. (\ref{diffusion}),
which can be interpreted as an advection of the magnetic field
lines by the motion of the electrons. In most astrophysical
contexts, this motion is shared by all the other particles,
corresponding to an ideal magneto-hydrodynamic motion (e.~g.,
\citealt{Kulsrud}). However, in neutron stars, other variants are
possible \citep{GR92}, such as a motion of all the charged
particles with respect to the neutrons (``ambipolar diffusion'',
e.~g., \citealt{Hoyos08}), or by only the most mobile charge
carrier (``Hall drift''; e.~g., \citealt{R07Hall}), each of which
has quite distinctive properties.

\section{Magnetic fields in neutron stars}

Many neutron stars are detected as pulsars, whose regular
pulsations in the radio, X-ray, and/or optical bands are produced
by a strong magnetic field turning around at the stellar rotation
period $P$. These periods slowly increase in time, i.~e., the
neutron stars lose rotational energy, probably through magnetic
coupling with their surroundings. Modelling this coupling as
electromagnetic radiation from a dipole rotating in vacuum,
oriented orthogonally to the rotation axis, one can infer the
surface magnetic field strength $B\propto\sqrt{P\dot P}$, where
$\dot P$ is the time-derivative of the rotation period (e.~g.,
\citealt{Shapiro}). Inferred fields range from $10^8~\mathrm{G}$
in millisecond pulsars up to $10^{15}~\mathrm{G}$ in soft
gamma-ray repeaters (SGRs), the latter being the strongest
magnetic fields known in the Universe, but still 3 orders of
magnitude weaker than the strongest that might conceivably be
present in neutron stars according to eq.~(\ref{Bmax}). In this
(dynamical) sense, neutron star magnetic fields are quite weak, as
they are in all other stars known so far. In spite of this, the
magnetic field may be the main agent breaking the axial symmetry
of the mass distribution in a rotating a neutron star, in this way
producing precession (Wasserman 2003), as appears to be observed
in some pulsars, and gravitational waves, which might quickly
reduce the rotation rate of newborn neutron stars (Cutler 2002).

In radio pulsars, the rotational energy loss can account for the
whole observed energy output (relativistic particles and
electromagnetic radiation). For the strongly magnetized, but
slowly rotating SGRs and anomalous X-ray pulsars (AXPs), however,
the observed X-ray luminosity is much larger than the rotational
energy loss rate, so an additional source of energy is required,
the most likely being the decay of their magnetic field
\citep{TD96}. This would make these objects be the only known
magnetically powered stars, or ``magnetars''. An interesting way
of probing the strong magnetic fields inside these objects appears
to be the quasi-periodic oscillations recently detected following
two large flares of SGRs and interpreted as crustal shearing modes
coupled to Alfv\'en waves travelling through the stellar core
\citep{Levin07}.

In very old neutron stars, such as millisecond pulsars and
low-mass X-ray binaries, the magnetic field is $<10^9\mathrm{G}$,
weaker than in young neutron stars, such as radio pulsars and
high-mass X-ray binaries ($\sim 10^{11-14}\mathrm{G}$), suggesting
that the magnetic field strength decays with time, perhaps induced
by accretion of matter from the binary companion (e.~g.,
\citealt{Payne07} and references therein). Magnetic field decay
within the population of single radio pulsars has also been
suggested by some authors \citep{Ostriker69,Narayan90} but
disputed by others \citep{Bhattacharya92,Faucher06}, and does not
seem to be well established.

\section{Origin of the magnetic field}

A natural hypothesis to explain the origin of the strong fields
observed in neutron stars is the compression of the magnetic flux
already present in the progenitor stars. It led \citet{Woltjer64}
to predict field strengths of $10^{14-16}~\mathrm{G}$, before any
neutron stars had been identified observationally. Many authors
(e.~g.,
\citealt{Ruderman72,R01b,Ferrario05a,Ferrario05b,Ferrario06}) have
pointed out that the distribution of magnetic fluxes is very
similar in magnetic A and B stars, white dwarfs, and neutron
stars, in this way providing support for the hypothesis of the
magnetic fluxes being generated on or even before the
main-sequence stage and then inherited by the compact remnants.

On the other hand, \citet{TD93} suggested that newborn neutron
stars are likely to combine vigorous convection and differential
rotation, making a dynamo process likely to operate in them. They
predicted fields up to $10^{15-16}~\mathrm{G}$ in neutron stars
with few-millisecond initial periods, and suggested that such
fields could explain much of the phenomenology associated with
SGRs and AXPs \citep{DT92,TD95,TD96}, some of which were later
confirmed to spin down at a rate consistent with a strong dipole
field ($10^{14-15}\mathrm{G}$; \citealt{Kouveliotou98,Woods99}).

Of course, the two processes are not mutually exclusive. A strong
field might be present in the collapsing star, but later be
deformed and perhaps amplified by some combination of convection,
differential rotation, and magnetic instabilities \citep{Tayler73,
Spruit02}. The relative importance of these ingredients depends on
the initial field strength and rotation rate of the star. For both
mechanisms, the field and its supporting currents are not likely
to be confined to the solid crust of the star, but distributed in
most of the stellar interior, which is mostly a fluid mixture of
neutrons, protons, electrons, and other, more exotic particles.

\section{Persistent, ordered field structures}

The magnetic fields of neutron stars, like those of upper main
sequence stars and white dwarfs, appear to be ordered (with a
roughly dipolar external configuration) and persistent (for much
longer than a solar cycle, perhaps for the entire existence of
these stars). As mentioned above, their magnetic flux
distributions are similar, which also implies that their ratios of
magnetic to gravitational energy (or magnetic stress to fluid
pressure) are similarly small in all of them. The Lorentz force is
much smaller than the pressure gradient and the gravitational
force that are dominant in establishing the hydrostatic
equilibrium in the star. Therefore, the hydro-magnetic equilibrium
state can be considered as a small perturbation to an
unmagnetized, ``background'' hydrostatic equilibrium (denoted by a
subscript ``0''), in which the (conceptual) introduction of the
magnetic field forces the fluid to displace from its ``initial''
position, $\vec r\to\vec r+\vec\xi(\vec r)$. This causes small
perturbations of density, which are customarily described in two
complementary ways:
\begin{itemize}
\item Eulerian perturbations, which compare the density at the
same point in space, before and after the perturbation,
\begin{equation}\label{Eulerian}
\delta\rho(\vec r)=\rho(\vec r)-\rho_0(\vec r), \end{equation}
\item Lagrangian perturbations, the change in the same fluid
element before and after being displaced,
\begin{equation}\label{Lagrangian}
\Delta\rho(\vec r)=\rho(\vec r+\vec\xi[\vec r])-\rho_0(\vec r).
\end{equation}
\end{itemize}
These two descriptions are related by
\begin{equation}\label{advect}
\Delta\rho-\delta\rho=\vec\xi\cdot\nabla\rho_0={d\rho_0\over
dr}\xi_r,
\end{equation}
and exactly the analogous relations hold for the Eulerian and
Lagrangian perturbations of the pressure, $\delta P$ and $\Delta
P$. These perturbations must satisfy the force balance condition
\begin{equation}\label{force}
{\vec j\times\vec B\over c}-\nabla\delta
P-\delta\rho~\nabla\psi_0=0,
\end{equation}
where $\psi_0(r)$ is the background gravitational potential
(assumed to be unperturbed, in the so-called ``Cowling
approximation'').

Another important, shared property of these stars is that a large
part (if not all) of their interior is stably stratified (i.~e.,
it resists convective overturn). In the case of upper
main-sequence envelopes and white dwarfs, this is because of a
radially increasing entropy; in the case of neutron stars, because
of a radial dependence in the fraction of protons, electrons, and
possibly other particles. This has the consequence that the
adiabatic sound speed, relating the Lagrangian pressure and
density perturbations of a given fluid element (that conserve
entropy and composition), $c_s^2=\Delta P/\Delta\rho$, is larger
than the background derivative
$dP_0/d\rho_0=(dP_0/dr)/(d\rho_0/dr)$, in which entropy or
composition are changing (e.~g., \citealt{RG92}). The Lagrangian
perturbations can be directly related to the divergence of the
fluid displacement field $\vec\xi(\vec r)$ causing them,
\begin{equation}\label{Lagrangian}
{\Delta P\over c_s^2}=\Delta\rho=-\rho_0\nabla\cdot\vec\xi.
\end{equation}
From all the above, one obtains that the Eulerian perturbations
$\delta\rho$ and $\delta P$ are linearly independent combinations
of $\nabla\cdot\vec\xi$ and $\xi_r$,
\begin{eqnarray}\label{linear}
\delta\rho=-\rho_0\nabla\cdot\xi-{d\rho_0\over dr}\xi_r,\\
\delta P=-\rho_0c_s^2\nabla\cdot\xi-{dP_0\over dr}\xi_r,
\end{eqnarray}
and can therefore be regarded as independent variables.

Thus, a given, sufficiently weak magnetic field $\vec B(\vec r)$
corresponds to an equilibrium configuration if two independent
scalar functions $\delta P$ and $\delta\rho$ can be found that
satisfy eq.~(\ref{force}). Since the latter is a 3-component
vector equation, this will not generally be possible. Therefore,
it imposes a condition on $\vec B$ that can be written as
\begin{equation}\label{condition}
\hat r\cdot\nabla\times[(\nabla\times\vec B)\times\vec B]=0,
\end{equation}
a single, scalar condition on the magnetic field. This condition
is much less restrictive than the often imposed force-free
condition, $(\nabla\times\vec B)\times\vec B=0$
\citep{Broderick08}, relevant for the opposite limit of
dynamically dominant fields, or even the condition
$\nabla\times[(\nabla\times\vec B)\times\vec B/\rho]=0$
\citep{Haskell07}, required only for barotropic fluids with a
unique pressure-density relation $\rho(P)$, which are not stably
stratified.

The assumption of axial symmetry, with
\begin{equation}\label{axisymmetry}
\vec
B=\nabla\times[\alpha(r,\theta)\nabla\phi]+\beta(r,\theta)\nabla\phi,
\end{equation}
where $\nabla\phi=\hat\phi/(r\sin\theta)$, considerably simplifies
the problem of constructing equilibria, since in this case
eq.~(\ref{force}) implies that the Lorentz force can have no
azimuthal component, $(\vec j\times\vec B)_\phi=0$, therefore
surfaces of constant $\beta$ coincide with those of constant
$\alpha$ (i.~e., $\beta=\beta[\alpha(r,\theta)]$). These
``allowed'' magnetic field configurations produce two independent
force components in the meridional plane, which can generally be
cancelled by an appropriate choice of the functions $\delta P$ and
$\delta\rho$, so no additional conditions are imposed on $\vec
B(\vec r)$ to correspond to an equilibrium.

Of course, in order to be viable, a certain magnetic field
configuration must correspond to a {\it stable} equilibrium, which
is much more difficult to characterize and still largely an open
problem. Analytic attempts to search for stable magnetic field
configurations have failed, only yielding the general result that
both purely toroidal fields $\vec B=\beta(r,\theta)\nabla\phi$ and
purely poloidal fields $\vec
B=\nabla\times[\alpha(r,\theta)\nabla\phi]$ are unstable
\citep{Tayler73,Flowers77}, and the speculation that linked
toroidal and poloidal fields might stabilize each other, yielding
a stable equilibrium \citep{Prendergast56,Wright73}. Recent MHD
simulations \citep{BS04,BS06,BN06} have shown initially complex,
``random'' magnetic fields to evolve on an Alfv\'en-like timescale
into a roughly axisymmetric, linked poloidal-toroidal
configuration that persisted for a resistive timescale and thus
might be a good approximation to the field structures in
upper-main sequence, white dwarf, and neutron stars.

Once a stable, ideal-MHD equilibrium magnetic field has been
established, it will survive for many Alfv\'en times, but not
forever, since there are several dissipative processes by which it
could evolve on long time scales
\citep{GR92,R07Potsdam,R07Hall,Hoyos08}, possibly matching the
times on which magnetar fields appear to decay \citep{TD96}.

\section{Conclusions}

Neutron stars are fascinating objects with extreme properties,
which include the strongest magnetic fields in the Universe.
Nevertheless, these share properties with those of other stars,
among these, that they are weak in the sense of producing only
small disturbances to the structure of the respective stars. Some
progress has been made in understanding possible magnetic field
configurations and their evolution, but there is still much left
to do.

\acknowledgements
      The author thanks H. Spruit and C. Thompson
      for many stimulating and informative conversations, and
      T. Akg\"un and J. Hoyos for a very careful reading that improved the
      quality of this manuscript.
      This work was supported by FONDECYT (Chile)
      Regular Research Grant 1060644.


\begin{thebibliography}



\bibitem[Baym et al.(1969)Baym, Pethick, \& Pines]{Baym69}  Baym, G., Pethick, C., \& Pines, D. 1969, Nature, 224,
674


\bibitem[Bhattacharya et al.(1992)]{Bhattacharya92} Bhattacharya,
D., Wijers, R.~A.~M.~J., Hartman, J.~W., \& Verbunt, F. 1992,
A\&A, 254, 198


\bibitem[Braithwaite \& Spruit(2004)]{BS04} Braithwaite, J., \& Spruit, H. 2004,
   Nature, 431, 819

\bibitem[Braithwaite \& Nordlund(2006)]{BN06} Braithwaite, J., \& Nordlund, \AA. 2006,
      A\&A, 450, 1077

\bibitem[Braithwaite \& Spruit(2006)]{BS06} Braithwaite, J., \& Spruit, H. 2006,
      A\&A, 450, 1097

\bibitem[Broderick \& Narayan(2008)]{Broderick08} Broderick,
A.~E., \& Narayan, R. 2008, \mnras, 383, 943


\bibitem[Cutler(2002)]{Cutler02} Cutler, C. 2002, Phys.~Rev.~D, 66, 084025



\bibitem[Duncan \& Thompson(1992)]{DT92} Duncan, R.~C., \&
   Thompson, C. 1992, ApJ, 392, L9

\bibitem[Faucher-Gigu\`ere \& Kaspi(2006)]{Faucher06}
Faucher-Gigu\`ere, C.-A., \& Kaspi, V.~M. 2006, \apj, 643, 332


\bibitem[Ferrario \&
   Wickramasinghe(2005a)]{Ferrario05a} Ferrario, L., \&
   Wickramasinghe, D.~T. 2005a, MNRAS, 356, 615

\bibitem[Ferrario \&
   Wickramasinghe(2005b)]{Ferrario05b} Ferrario, L., \&
   Wickramasinghe, D.~T. 2005b, MNRAS, 356, 1576

\bibitem[Ferrario \&
   Wickramasinghe(2006)]{Ferrario06} Ferrario, L., \&
   Wickramasinghe, D.~T. 2006, MNRAS, 367, 1323


\bibitem[Flowers \& Ruderman(1977)]{Flowers77} Flowers, E.,
   \& Ruderman, M.~A. 1977, ApJ, 215, 302


\bibitem[Glendenning(2000)]{Glendenning} Glendenning, N. K. 2000,
Compact Stars: Nuclear Physics, Particle Physics, and General
Relativity, 2nd edition, New York: Springer

\bibitem[Goldreich \& Reisenegger(1992)]{GR92} Goldreich, P., \& Reisenegger, A. 1992,
      ApJ, 395, 250

\bibitem[Haensel et al.(2007)Haensel, Potekhin, \& Yakovlev]{Haensel} Haensel, P., Potekhin,
A.~Y., \& Yakovlev, D.~G. 2007, Neutron Stars 1: Equation of State
and Structure, New York: Springer

\bibitem[Haskell et al.(2007) Haskell, Samuelsson, Glampedakis,
   \& Andersson]{Haskell07} Haskell, B., Samuelsson, L.,
   Glampedakis, K., \& Andersson, N. 2007, preprint
   (arXiv:0705.1780v1[astro-ph])

\bibitem[Hoyos et al.(2008)Hoyos, Reisenegger, \& Valdivia]{Hoyos08} Hoyos,
J., Reisenegger, A., \& Valdivia, J.~A. 2008, A\&A, submitted
(arXiv:0801.4372v1[astro-ph]))




   \bibitem[Kouveliotou et al.(1998)]{Kouveliotou98} Kouveliotou,
   C., et al. 1998, Nature, 393, 235

   \bibitem[Kulsrud(2005)]{Kulsrud} Kulsrud, R.~M. 2005, Plasma
   Physics for Astrophysics, Princeton University Press


   \bibitem[Levin(2007)]{Levin07} Levin, Y. 2007, MNRAS, 377, 159

   \bibitem[Lyne et al.(2004)]{Lyne} Lyne, A.~G., et al. 2004,
   Science, 303, 1153









   \bibitem[Narayan \& Ostriker(1990)]{Narayan90} Narayan, R., \& Ostriker, J.~P.
   1990, ApJ, 352, 222

   \bibitem[Ostriker \& Gunn(1969)]{Ostriker69} Ostriker, J.~P.,
   \& Gunn, J.~E. 1969, ApJ, 157, 1395


   \bibitem[Payne \& Melatos(2007)]{Payne07} Payne, D.~J.~B. \&
   Melatos, A. 2007, MNRAS, 376, 609

   \bibitem[Pethick(1992)]{Pethick92} Pethick, C.~J. 1992, in The
   Structure and Evolution of Neutron Stars, D. Pines, R. Tamagaki, \& S. Tsuruta, eds.,
   p. 115

\bibitem[Pons \& Geppert(2007)]{Pons} Pons, J., \& Geppert, U. 2007,
A\&A, 470, 303


\bibitem[Prendergast(1956)]{Prendergast56} Prendergast, K.~H. 1956, ApJ, 123, 498



   \bibitem[Reisenegger(2001a)]{R01a} Reisenegger, A. 2001a, ApJ,
   550, 860

   \bibitem[Reisenegger(2001b)]{R01b} Reisenegger, A. 2001b, in
   Magnetic Fields across the Hertzsprung-Russell Diagram, ASP
   Conference Series, vol. 248, eds. G. Mathys, S.~K. Solanki, \& D.~T. Wickramasinghe,
   p. 469


   \bibitem[Reisenegger(2007)]{R07Potsdam} Reisenegger, A. 2007,
   AN, 328, 1173


   \bibitem[Reisenegger et al.(2006)]{R06} Reisenegger, A.,
   Jofr\'e, P., Fern\'andez, R., \& Kantor, E. 2006, ApJ, 653,
   568

   \bibitem[Reisenegger et al.(2007)]{R07Hall} Reisenegger, A., Benguria, R.,
   Prieto, J.~P., Araya, P.~A., \& Lai, D. 2007, A\&A, 472, 233

   \bibitem[Reisenegger \& Goldreich(1992)]{RG92} Reisenegger, A.,
   \& Goldreich, P. 1992, ApJ, 395, 240


   \bibitem[Ruderman(1972)]{Ruderman72} Ruderman, M. 1972, \araa,
   10, 427

   \bibitem[Sagert et al.(2006)]{undergraduates2} Sagert, I.,
   Hempel, M., Greiner, C., \& Schaffner-Bielich, J. 2006, Eur. J.
   Phys. 27, 577


   \bibitem[Shapiro \& Teukolsky(1983)]{Shapiro} Shapiro, S. L., \&
   Teukolsky, S. A. 1983, Black Holes, White Dwarfs, and Neutron
   Stars (New York: Wiley)

   \bibitem[Silbar \& Reddy(2004)]{undergraduates1} Silbar, R.~R.,
   \& Reddy, S. 2004, Am. J. Phys., 72, 892; erratum 2005, Am. J.
   Phys., 73, 286

   \bibitem[Spruit(2002)]{Spruit02} Spruit, H. 2002, A\&A, 381,
   923


   \bibitem[Tayler(1973)]{Tayler73} Tayler, R.~J. 1973, MNRAS, 161,
   365

   \bibitem[Thompson \& Duncan(1993)]{TD93} Thompson, C., \& Duncan, R. 1993, ApJ, 408, 194

   \bibitem[Thompson \& Duncan(1995)]{TD95} Thompson, C., \& Duncan, R. 1995, MNRAS, 275, 255

   \bibitem[Thompson \& Duncan(1996)]{TD96} Thompson, C., \&
   Duncan, R.~C. 1996, ApJ, 473, 322


   \bibitem[Thorne(1994)]{Thorne} Thorne, K.~S. 1994, Black Holes and Time Warps:
   Einstein's Outrageous Legacy, New York: Norton



\bibitem[Wasserman(2003)]{Wasserman03} Wasserman, I. 2003, MNRAS,
341, 1020



\bibitem[Woltjer(1964)]{Woltjer64} Woltjer, L. 1964, \apj, 140,
1309

\bibitem[Woods et al.(1999)]{Woods99} Woods, P.~M. 1999, ApJ,
524, L55

\bibitem[Wright(1973)]{Wright73} Wright, G.~A.~E. 1973, MNRAS, 162,
339

\bibitem[Yakovlev et al.(2001)]{Yakovlev01} Yakovlev, D. G., et al. 2001, Phys.~Rep., 354, 1



\end{thebibliography}
\end{document}